# Impact detections of temporarily captured natural satellites

David L. Clark[1,2,3], Pavel Spurný[4], Paul Wiegert[2,3], Peter Brown[2,3], Jiří Borovička[4], Ed Tagliaferri[5], Lukáš Shrbený[4]

[1] Department of Earth Sciences, University of Western Ontario, London, ON N6A 5B8, Canada
[2] Department of Physics and Astronomy, University of Western Ontario, London, ON N6A 5B8, Canada
[3] Centre for Planetary Science and Exploration, University of Western Ontario, London, Ontario N6A 5B8, Canada
[4] Astronomical Institute, Academy of Sciences of the Czech Republic, CZ-251 65 Ondřejov, Czech Republic.
[5] ET Space Systems, 5990 Worth Way, Camarillo, California 93012, USA
Correspondence author email: dclark56@uwo.ca
Short Title: Impact detections of natural satellites.

**Abstract:** Temporarily Captured Orbiters (TCOs) are Near-Earth Objects (NEOs) which make a few orbits of Earth before returning to heliocentric orbits. Only one TCO has been observed to date, 2006 $RH_{120}$, captured by Earth for one year before escaping. Detailed modeling predicts capture should occur from the NEO population predominantly through the Sun-Earth L1 and L2 points, with 1% of TCOs impacting Earth and approximately 0.1% of meteoroids being TCOs. Although thousands of meteoroid orbits have been measured, none until now have conclusively exhibited TCO behaviour, largely due to difficulties in measuring initial meteoroid speed with sufficient precision. We report on a precise meteor observation of January 13, 2014 by a new generation of all-sky fireball digital camera systems operated in the Czech Republic as part of the European Fireball Network, providing the lowest natural object entry speed observed in decades long monitoring by networks world-wide. Modeling atmospheric deceleration and fragmentation yields an initial mass of ~5 kg and diameter of 15 cm, with a maximum Earth-relative velocity just over 11.0 km/s. Spectral observations prove its natural origin. Back-integration across observational uncertainties yields a 92 - 98% probability of TCO behaviour, with close lunar dynamical interaction. The capture duration varies across observational uncertainties from 48 days to 5+ years. We also report on two low-speed impacts recorded by US Government sensors, and we examine Prairie Network event PN39078 from 1965 having an extremely low entry speed of 10.9 km/s. In these cases uncertainties in measurement and origin make TCO designation uncertain.

# 1   Introduction

The existence of a population of temporarily captured terrestrial natural satellites (also called temporarily captured orbiters (TCOs) or minimoons) was first suggested almost 100 years ago (Chant, 1913) (Denning, 1916). Typically asteroids which pass near Earth may have their paths altered by the Earth's gravitational attraction, but only a few might be captured.  Even that is only a short-term state, which lasts typically only a few orbits of the TCO around the Earth. TCOs were observationally confirmed with the discovery of 2006 RH$_{120}$ (Kwiatkowski, et al., 2009). This ~5m asteroid was captured for nearly one year in mid-2006 before returning to an unbound state.

TCOs are of interest because they spend a relatively long time in orbits which are very accessible from Low-Earth Orbit because of their low Delta-Vs, and hence are easy for spacecraft to visit, as with the proposed Asteroid Retrieval Mission (ARM) (Mazanek, Brophy, & Merrill, 2013). Moreover, their extended proximity to Earth allows detailed remote-sensing to characterize the body in-situ.  The low entry velocity of TCOs also results in a high meteorite survival fraction offering the possibility of both in-space observation and material recovery. The steady-state TCO population as a function of size is sensitively dependent on the size frequency distribution (SFD) of the near Earth asteroid (NEA) population and measurement of the TCO SFD would place constraints on the true NEA SFD at small sizes.

While the majority of TCO's escape back into interplanetary space, a significant proportion are expected to collide with our planet as meteors. Based on dynamical simulations of the TCO population, Granvik et al. (2012) estimated that approximately 0.1% of meteoroids impacting Earth should have been TCOs prior to impact. Moreover, long-lived TCOs have a much higher (~20%) probability of Earth impact. Yet, we are unaware of any confirmed TCO in the population of Earth-impacting meteors, until now.  Bolin et al (2014) examined contemporary small meteor video detection systems and potential TCO detection rates, noting the difficulty in detecting fainter meteoroids impacting at low TCO speeds and in particular the challenge of getting precise enough metric solutions to confirm an event as a TCO.

In the early morning of January 13 2014 at 03:01:38 UT a 5 kg object entered Earth's atmosphere in an approximately 33° sloped and 77 km long southerly trajectory over the border areas of Czech Republic, Germany and Austria. Detected by new high precision digital camera systems of the Czech portion of the European Fireball Network (or EN), EN130114 exhibited the lowest initial velocity of any natural object ever observed by the network (Table 1). The EN observation is notable for its high precision which translates directly into a narrow possible range for its initial speed. As discussed in the next sections, based on the deceleration corrected velocity at initial atmospheric contact, this meteoroid was almost certainly a TCO prior to impact. Remarkably, spectra were also secured of the event, confirming it as a natural (as opposed to man-made) object.

Another dataset we examine are US Government sensor detections of meter-sized and larger objects impacting the atmosphere, which have begun to be regularly disseminated via the NASA Jet Propulsion Laboratory's Fireball and Bolide Reports Web page http://neo.jpl.nasa.gov/fireballs (Yeomans & Baalke, 2014). In some cases the atmospheric contact position and velocity states provided are sufficient to permit reconstruction of pre-contact trajectory and heliocentric orbit. Among the impacts listed is an event occurring on 2014-06-26 05:54:41 UT over northeast Antarctica (near the Indian Ocean) which released 0.2 kiloton of TNT equivalent energy. The pre-impact geocentric velocity is noted as 11.2 km/s. This event may also be a TCO within the precision of the state vector provided, though at a lower probability than EN130114. A second event in the Fireball and Bolide reports, a 0.12 kiloton of TNT event on 2008-07-01 over the California-Nevada border, is recorded with a pre-impact velocity of only 9.8 km/s, an impossibly low velocity for a natural object.

Finally we examine an intriguing Prairie Network (PN) fireball PN39078 which occurred on November 14, 1965 (McCrosky, Shao, & Posen, 1978). PN39078 had an uncommonly low measured velocity of 10.88 km/s. Unfortunately, the original records are not readily available for the quantification of the uncertainties of measurement, uncertainties being the critical factor in making any statement on TCO likelihood.

In what follows we examine these four cases in detail with particular attention to the corrections required for atmospheric deceleration. Our goal is to establish whether any may have been TCOs prior to impact and the implication for the TCO steady-state population and the NEA SFD more generally. Our work builds on a similar analysis by Bolin et al (2014) which focussed on TCO meteor detection at smaller sizes.

## 2 Observations

### 2.1 EN130114

EN130114 was detected by two cameras of the Czech component of the European Fireball Network (Spurný, Borovička, & Shrbený, 2007) (Figure 1). The Czech portion of the EN has been upgraded as of 2014 to high resolution fully digital cameras, increasing fireball detection capability with improved astrometric and photometric fidelity on fireballs captured. Table 1 summarizes the in-atmosphere trajectory, mass and fireball type determined for EN130114.

Figure 2 and Figure 3 show the astrometric residuals and light curve as a function of time from these camera reductions. The first measured velocity at 64 km altitude was 10.90 ± 0.04 km/s. Based on the observed end height, the fireball is type I, consistent with a chondritic-type meteoroid (Ceplecha & McCrosky, 1997). Deceleration modelling to fit the observed trajectory and light curve yields a 5 kg mass and ~0.15 m diameter for the initial meteoroid. Small meteorites up to a few hundreds of grams probably reached the ground, but no systematic search has been conducted.

Spectra acquired during the event (Figure 4) preclude the object being man-made. The dominant emission is by neutral Na. Other visible lines belong to neutral Fe and Cr. These emissions are common in natural meteoroids (Vojáček, Borovička, Koten, Spurný, & Štork, 2015) (Borovička J., 1994) and the spectrum is fully consistent with a natural object. The Mg line was faint, which can be ascribed to the very low velocity of the fireball. Artificial bodies exhibit different spectra. The Hayabusa spacecraft contained exotic lines of Cu and Mo in the same spectral range (Abe, et al., 2011). The near-UV spectra obtained during the ESA ATV1 re-entry show dominance of Al and weakness of Fe (Löhle, Wernitz, Herdrich, Fertig, Röser, & Ritter, 2011).

## 2.2 JPL20140626

Table 2 provides the speed, local radiant, energy and height of peak brightness for our second possible TCO candidate JPL20140626. The object's measured speed of 11.2 km/s corresponds to an approximate 13,000 kg object of 2m diameter assuming a chondritic bulk density. We assume that the speed given is the average across the entire visible path of the US Government sensors. Following a similar event reported by Klekociuk et al. (2005) where a meteoroid of comparable speed and size was first detected at ~70 km altitude, we also assume the speed average corresponds approximately to the height interval 70 – 28.5 km. Fragmentation will increase the deceleration of the main body; in the absence of any further ablation information on this object we can model only a simple lower limit to the deceleration assuming single body ablation.

## 2.3 JPL20080701

JPL20080701 is recorded as having an initial energy of 0.12 kiloton of TNT and occurred over the California-Nevada border west of Las Vegas. The extremely low speed of 9.8 km/s, if accurate, would equate to a mass of 10,000 kg and diameter of 2m, again assuming chondritic bulk density. The peak brightness altitude of 36.1 km suggests it experienced less deceleration at play that than JPL20140626, but that similar methods for modelling deceleration could be used. However, the low velocity is problematic, in that such a velocity cannot be attributed to a natural object. Clearly deceleration and measurement error need further consideration.

## 2.4  PN39078

PN39078 was an estimated 5kg object recorded at JD 2439078.796 (November 14, 1965 07h06m ± 2m UT) by the Prairie Network (McCrosky, Shao, & Posen, 1978).  The event was recorded by two stations: 12E and 9W with an approximate location of 99.8° W and 41.65°N. Positional data was confirmed along with radiant and trajectory information extracted from the PNSAO.DAT dataset documented in Ceplecha & McCrosky (1997). The measured speed 10.8 km/s at altitude 65 km is similar to EN 130114. We note that a handwritten remark on the original data files (examined in 1993) noted the event as a possible re-entry, without further details.

To establish TCO classification, the initial speed prior to atmospheric deceleration must be determined. It is apparent that the observed in-atmosphere speeds for these TCO candidates are well below that expected for an object on an unbound geocentric orbit. However, at their initial detection depth in the atmosphere some prior deceleration is expected. In the next section we apply an entry model to EN130114, and to a less rigorous degree the JPL events, and estimate the probable range of initial speed prior to significant deceleration.

## 3  Methods

### 3.1  Estimating Exoatmospheric initial Conditions

The slow entry speed of TCO candidates necessitates the careful consideration of atmospheric deceleration prior to initial detection.  Deceleration models provide the mapping between observations and initial exoatmospheric heliocentric position and velocity states from which gravitation-only back integrations may be performed.  The models described below take into account both deceleration due to atmospheric drag and acceleration due to gravity during the bolide phase.

The EN130114 event shows remarkable deceleration from the beginning of velocity measurements at altitude 64 km. The deceleration could be fitted by a single body model, i. e.

with no fragmentation. Assuming $\Gamma A=1.0$ ($\Gamma$ is the drag coefficient and A is the shape coefficient) and meteoroid density 3000 kg/m$^3$, the resulting initial mass was 1.2 kg. The ablation coefficient was quite high in this solution, 0.087 s$^2$/km$^2$. Such a model, however, does not explain the observed light curve (Figure 3). The light curve shows the slope expected for a single body at altitudes 63-60 km (the steeper increase before that may be due to onset of ablation) but a steep increase in luminosity between 60 and 50 km. The maximum magnitude of -7.5 is reached at 50 km. To explain the luminosity increase, several fragmentation events between 60 and 50 km are needed.

Three fragmentation models for EN130114 are considered: Models 1 and 2 representing a 3 g/cm$^3$ 5 kg meteoroid at extreme ends of the possible initial velocities. An ablation coefficient of 0.005 s$^2$/km$^2$ is assumed along with the luminous efficiency function following the values used for an analysis of the Košice event (Borovička, et al., 2013). These were considered a priori as the best models, but the curious nature of the event led us to consider an unlikely 7 kg 1 g/cm$^3$ cometary density corresponding to the highest possible initial velocity (low density leads to stronger deceleration). Figure 3 shows the resulting modelled light curves compared to that observed. The two 3 g/cm$^3$ and one 1 g/cm$^3$ fragmentation models all match the observed light curve, at least the ascending part, which is important for this study, and are therefore considered for further dynamical analysis.

For both the high density and low density models, an object's direction of travel relative to the Earth (radiant) along with a speed are calculated at intervals of 0.02sec. The object is assumed to move along a purely linear trajectory during light production. Though the trajectory is curved due to the effect of gravity, this effect is small (deflection of ~ 0.2°) and difficult to measure due to light curve flares and variability, interference from clouds, etc. We allow for this small curvature by increasing the uncertainty in the radiant direction by 0.2° in the vertical direction, which is primarily in the declination of the radiant, as the bolide's azimuth is 343°, just west of north. The exoatmospheric initial conditions are extracted at an altitude just above where atmospheric drag becomes important (94 km), and the motion of the body prior to this point is assumed to be affected only by gravitational interactions. Models 1 and 2 (speeds at 94 km 10.90 and 10.93 km/s, respectively) were combined into a single high density model covering the velocity range of the two, and Model 3 (11.02 km/s) is carried forward as a low-likelihood low density model. Note that the speed of the no-fragmentation solution was close to Model 2 (10.94 km/s). All given speeds are relative to the Earth's surface. Figure 5 plots the estimated in atmosphere velocities (relative to Earth's center, i.e. corrected to Earth's rotation) and their corresponding spread due to measurement Gaussian uncertainties (3σ used for illustration purposes only, 1σ in later analysis) by altitude for both models. Gravitational acceleration dominates at high altitudes, with atmospheric drag dominating at lower altitudes producing deceleration.

PN39078's atmospheric deceleration was modelled in the same fashion as the EN130114 scenario. In this case we have no radiometric light curve, only photographic. The maximum magnitude was -5 and the meteoroid was likely of about 1 kg. The fragmentation seems to be not so severe and all models are within 10.90 ± 0.02 km/s at 94 km, using data from station 12E. Station 9W, however, gives speed by 0.05 km/s lower, which suggest a possible problem with the geometric solution (radiant position). Bad geometry would also affect speed values.

For JPL20140626 we applied the FM model of Ceplecha and ReVelle (2005) and mean parameters for the apparent ablation coefficient and shape-density factor (Ceplecha, et al., 1998) for chondritic and carbonaceous chondrite bodies and estimated the deceleration in the extreme case of no fragmentation. We find that the average minimum velocity decrease varies between 0.05 – 0.2 km/s by the peak brightness height of 28.5 km, noting that the true deceleration is likely higher. To evaluate the impact of drag on the TCO nature of JPL20140626, gravitational back-integration (see below) is performed four times with differing initial speeds using an unchanged velocity vector direction. Velocities used are the quoted velocity ($v_0$ km/s), $v_0 + 0.1$, $v_0 + 0.2$, and $v_0 + 0.3$ km/s. We use and quote results from a relatively small number of clones (1,000), as initial results indicate that further analysis of a larger number of clones would yield little additional information of value. We perform a fifth integration using an artificially reduced uncertainty on the contact velocity to both focus on theoretical TCO behaviour and to illustrate the need for high precision velocity measurements.

[8]

The JPL20080701 event is more problematic with the extremely low reported initial velocity of 9.8 km/s. Speeds significantly lower than Earth's escape velocity correspond to Earth launched non-natural objects, and in the case of the speed and trajectory of JPL20080701, a Western Pacific launch 3 to 5 hours prior to observation. But, the peak brightness altitude of 36.1 km is very high for a slow moving man-made object. Brown et al. (2015) comments on potential issues with accuracy based on comparison between ground-based measurements and some events in the JPL list. For the purposes of quantifying the deceleration and/or data correction required to achieve a non-zero probability of being a natural object, the calculated radiant was used, with the velocity magnitude increased in steps of 0.2 km/s until unbound and TCO behaviour became evident in back integrations (described below) with an increase of 1.2 km/s (11.0 km/s) and 1.4 km/s (11.2 km/s).

## 3.2 Back Integration of Probability Clones

The methods used for in-atmosphere and exoatmospheric back integration and orbit calculation are derived from those described in Clark (2010) in the ongoing search for serendipitous sky survey images of pre-impact meteoroids, and Clark and Wiegert (2011) in the numerical verification of Ceplecha's analytic meteoroid orbit determination method. We calculate possible meteoroid trajectories and orbital evolution by selecting 1,000-20,000 (depending on event and drag model) clones. These were generated by selecting initial conditions offset from the nominal fireball trajectory by a random amount drawn from a Gaussian distribution with a standard deviation set by the measurement uncertainties in each of the three radiant-velocity (R.A., Dec, and speed; or velocity Cartesian coordinates) and the speed, as well as the longitude, latitude and altitude above the WGS84 geoid. These clones each represent a possible state for the object, a state which is different from the nominal solution but which cannot be differentiated from it observationally due to measurement uncertainty. If the clones show consistent behaviour, then we can conclude that the real object showed the same behaviour; if not, then our observations are simply not precise enough to make a clear statement about the object in question.

We work directly from the measured or published quantities instead of derived ones such as orbital elements to minimize the effects of correlation. We assume the measured quantities are not significantly correlated. The generated clones are integrated backwards independently for a period of 5 years, their spread as one moves back in time providing as rigorous a measure of the backwards evolution of the orbital uncertainties as is available. For EN130114, the standard deviations of these distributions are taken from the calculated uncertainties from observations shown in Table 1. For JPL20140626 and JPL20080701, we assume the precision is set by the last significant digit of the JPL-provided value as shown in Table 2. Quoted geocentric Earth-fixed reference frame velocities were adjusted for both Earth rotation and Earth orbital motion. Similarly, for PN39078, where measurement uncertainties are not published, standard deviations estimated from expected measurement uncertainties of the detection system are used. In all three cases, these likely represent underestimates of the true uncertainties. For all events, the Everhart (1985) RADAU-15 $15^{th}$ order differential equation integrator is used to calculate the gravitational influences of the Sun, Earth, Moon, and major planets on each clone. Earth's J2 (2nd degree harmonic accounting for ellipsoidal shape) is considered, while post-Newtonian forces are ignored. Table 3 lists the initial conditions and uncertainties for all event integrations.

### 3.3   Clone Classification and Orbit Counting

Back-integration of event clones yield different behaviours as each has a slightly different initial position and velocity. We identify three broad classes of behaviour: 1) TCO behaviour, where an object orbits the Earth (see below); 2) 'unbound' clones which were never orbited by our planet prior to impact; and 3) clones which intersect the Earth's surface in the past. This last class, which we term 'sputniks', are physically impossible for natural objects as they would have had to be ejected from the Earth's surface in the recent past. Such a state is possible for man-made spacecraft only. We then subdivide these broad classes into six classes of behaviour seen in the backwards simulation as TCO (no lunar influence), TCOL (lunar influenced TCO), UNB (unbound - clones which never orbited our planet prior to impact), UNBL (unbound with lunar influence), Sp0 (zero-orbit sputniks), and Spn (orbiting sputniks).  See Table 4 for complete descriptions.

The counting of the number of bound orbits of the Earth a clone experiences is non-trivial due to the varied and changing orientations of the clone trajectories.  EN130114 and PN39078 exemplify the complexity with initial high inclination prograde trajectories, with some clones evolving (in backward time) to low inclination retrograde trajectories. The number of orbits reported in this work is determined by recording transitions in the Earth-object distance derivatives (transitions from approaching to receding) while the object remains in the Earth environment (5 Hill Sphere radii).   Orbit counts determined in this way are independent of reference frame.  Trial comparisons with the cycle count approach of Granvik et al. (2012) confirm that the methods report very similar TCO behaviour statistics.

## 3.4  TCO Probability

We define TCO probability as the number of TCO clones divided by the total number of possible clones $P_{TCO} = \frac{N_{TCO} + N_{TCOL}}{N_{pos}}$ where $N_{pos} = N_{TCO} + N_{TCOL} + N_{UNB} + N_{UNBL}$ (or equivalently $N_{pos} = N_{Clones} - N_{Sp0} - N_{Spn}$). We introduce 'possible clones' in the denominator instead of the total number of clones studied because we consider that sputniks, the clones which must have been ejected from the Earth itself, are impossible. Though mathematically allowed, they are an artifact of the measurement uncertainties; they do not represent physically realistic cases and so are not included in our final statistics.

Initial back-integrations of EN130114's probability clone cloud revealed an unexpected complexity in the storage and analysis of clone ephemerides, resulting in the needs for bracketing of TCO probability results.  We see in Figure 6 two important attributes of the EN130114 cloud: a highly elliptical orbit with a very close Earth perigee, and a very quick spread of mean anomaly around that orbit.  The frequent clone close-approaches to Earth necessitate very high resolution (approximately one minute) integration steps for both force calculations and Sputnik detection. The RADAU integrator naturally adjusts internal step sizes as required, even if the resulting ephemeris is not recorded. However, to accurately record Sputnik scenarios, Earth intersection must be tested at high resolution on output from the integrator. Earth intersection detection within RADAU's force calculators is not accurate, in that RADAU is a predictive-corrective integrator, frequently performing force calculations on trial positions that are not real. We attempted to address this issue by performing the needed order 1,000,000 high resolution clone close approach integrations over the first year prior to contact. The resulting data processing overhead of maintaining this large number of ephemerides over random epochs proved too daunting for our current software architecture; the resolution to which we leave for future work. Instead, we bracket our probability results with pessimistically low results taken by sampling from the RADAU force calculations, and optimistically high results taken from lower resolution (1-hour) integrations which are known to miss some Sputnik scenarios.

[11]

## 4   Results

### 4.1   EN130114 Results

The back-integration of both high and low density models reveals TCO behaviour in all cases (See Table 5). From the simulation results, we find the preferred high density model exhibits near certain (92.1 - 98.6%) TCO probability, while the lesser low density model demonstrates 22.7 – 23.9% probability. In Figure 7 we plot, for each model, durations for which clones are in bound orbits within the Earth Hill Sphere by initial geocentric velocity, corresponding to the optimistic TCO probabilities. Figure 8 is a corresponding plot showing the number of TCO orbits against initial velocity. Impossible Sp0 (black) and Spn (blue) clones are plotted to better demonstrate velocity regimes; these Sputniks are ignored in later analysis. A pattern appears evident across models, listed from lowest to highest velocity: (a) a very low velocity band of impossible Sp0 clones evident in the high density models, and surmised as being possible for a lower velocity low density object, (b) a low velocity band of intermixed TCO, TCOL, and SPn clones prominent in the high density model and hinted at in the low density; (c) a mid-velocity band of impossible Sp0 clones evident in both models; (d) a band of high-velocity interspersed TCO, TCOL, and Spn clones prominent in the low density model with some representation in the high density, and (e) a band of higher velocity UNBs seen in both models but poorly represented in the high density model. Of particular note are the relatively narrow velocity ranges which exhibit TCO and TCOL behaviour. The 0.02 – 0.04 km/s width of these ranges underscores the measurement precision required for TCO determination. We performed a correlation analysis comparing clone behaviours against all possible pairings of the 6 clone initial position and velocity state coordinates. We found no correlations, demonstrating that clone behaviour is dependent strictly on velocity.

The lower velocity band of TCOs and Sputniks, most evident in the high density model, represents the scenario where the clones are in an elongated orbit with near-vertical inclination to the lunar orbital plane, with perigee near Earth and apogee near the Moon (see Figure 9). These clones, moving at low speeds at apogee, are highly likely to be perturbed by lunar passage (see Figure 6) explaining the dominance of TCOL clones (orange) over TCO clones (red) in Figure 7. In a second analysis performed to quantify the impact the somewhat arbitrary usage of one lunar Hill Sphere radius for clone categorization, a radius of two lunar Hill Spheres was used. We found In this case that TCOLs vastly outnumbered TCO clones, leaving only a few long lived TCO's with no lunar interactions remaining, all of which were still in the Earth's Hill Sphere at the end of the 5 year integration. This suggests that for the slow lunar apogee clones, the Moon plays a role in TCOs impacting the Earth. The compelling evidence for lunar influence can be seen in Table 5 and Figure 10; of the 2569 lunar interacting TCO clones, 3373 total TCO clones, or 3422 physically possible clones, nearly 2000 pass through the lunar Hill Sphere immediately before impacting Earth. This is strong evidence that the Moon played a significant role in the impact.

The higher velocity band of TCOs and Sputniks, representing almost all TCOs in the low density model, is comprised of clones whose orbits initially extend 2-4 lunar orbit radii (approximately 1/2 – 1 Earth Hill Sphere radii). Under the increased solar gravitational influence, the initially high-inclination higher velocity orbits broaden, decline in inclination, increase in perigee distance from the Earth, and take on a retrograde Earth-orbit direction (See panels b) and c) of Figure 11). The broad clone orbits and increased influence of the Sun result in the largest proportion of clones being pure TCO (red in the plots) dominating the TCOL clones. The TCOL population exists due to the fact the reduced inclination clone orbits now make clones candidates for lunar focussing on both inbound and outbound orbit legs. However, the actual influence of passing through the lunar Hill Sphere is less than with low velocity clones near apogee, as the Moon-relative velocities of the clones are much greater.

[13]

Both low and high velocity bands exhibit significant variation among clones in the duration of TCO behaviour (Figure 12). The low velocity TCO clones remain in the Earth Hill Sphere a minimum of 50 days, while the high velocity clones remain a minimum of 48 days. Both velocity bands include a small number of clones (approximately 5-8% of all TCO clones) which continue to orbit the Earth the full 5-year integration period prior to impact. Figure 12 shows a significant portion of TCO's remain such for 250-500 days.

## 4.2 JPL20140626 Results

Table 5 summarizes the clone behaviour for 5 scenarios: the quoted velocity being the atmospheric contact velocity, incorporating 0.1, 0.2 and 0.3 km/s deceleration prior to observation, and the quoted velocity with an artificially small uncertainty range (0.01 km/s) similar to that of EN130114 observation. Figure 13 plots clone durations within the Earth Hill Sphere by initial geocentric velocity as done in Figure 7 for EN130114. Using the observed velocity, JPL20140626 has but a 10.5% probability of being a TCO. Almost 54% of the 1000 clones are gravitationally unbounded to the Earth, while nearly 40% are un-real Sp0 examples. Relatively few clones are left for TCO, TCOL and Spn behaviour. With the velocity range for TCO behaviour being small, the velocity uncertainty range is dominated by non-TCO candidates. The quoted object speed is a lower limit on contact speed. When incorporating decelerations of 0.1, 0.2, and 0.3 km/s, the probability of the object being a TCO is 2.6%, 0.6% and 0.0% respectively.

We repeat the analysis using artificially reduced standard deviations on the contact position, velocity and time inputs to get a picture of how results cluster around the mean object trajectory. The mean velocity being 11.24 km/s (again, to artificial precision) corresponds to UNB behaviours, with TCO, TCOL and Spn behaviour restricted to a narrow 0.015 km/s wide range. With the mean being relatively close to this range, the TCO probability increases markedly to 29%. The relatively low lunar proximity TCOL count as compared to TCO count shows similarity to the high velocity TCO clones described under EN130114 results. Detailed animation of the clone cloud confirms this, with TCO and TCOL clone back integration showing similar outer Earth Hill Sphere and low inclination dynamics as in Figure 11 for the EN130114 low density model.

### 4.3 JPL20080701 Results

The combination of the unknown and presumably wide uncertainties of the JPL list events combined with the unnaturally low entry speed of JPL20080701 make any TCO determination impossible. By increasing the calculated velocity in order to compensate for deceleration and measurement error, we found that an increase of 1.2-1.4 km/s was required for JPL20080701 to have a TCO or unbound orbit and hence be a natural object. With this increase, as with JPL20140626, the wide uncertainties enveloped a large range of behaviours: unbound, Sputnik, and a narrow band of TCO behaviour (See Figure 14).

### 4.4 PN39078 Results

As with the JPL events, the assumed uncertainties around measured and calculated atmospheric contact state parameters yield a wide variety of possible behaviours: from impossible Sputnik scenarios, through TCOs, to unbounded scenarios (See Table 5 and Figure 14). The extreme dominance of impossible Sputniks places doubt on the original measurements. The calculated TCO probability of 39% is questionable as the number of TCO's and unbounded clones from which it is calculated is very small. Of interest however is the dynamics of the TCO and Spn clones, which are in highly inclined orbits, many of which have an approximate lunar distance apogee similar to EN130114 (Figure 15). Unlike EN130114, none of the few TCO clones modelled enter the lunar Hill Sphere.

# 5  Discussion

The large number of EN reductions of fireballs (which we have estimated to be of order 1000) and the ever-increasing number of JPL reported events provide the opportunity to derive TCO frequency statements. Assuming the proposed 1% of all impacting meteoroids being TCOs, there should be of order 10 observations of TCO candidates from EN, which we do not see. However, the increased measurement accuracy possible from the new digital systems like the augmentation undertaken to the Czech portion of the EN makes future TCO observations more probable as measurement errors are lower than in earlier EN systems. The JPL list of events with measured velocities is not yet sufficiently large, nor of sufficient precision, to support TCO frequency statements.

The modelling of EN130114 highlights a TCO scenario of TCO behaviour where an object may enter into a relatively low energy orbit (near lunar distance apogee) and remain there for a significant period of time. It is interesting that of the four TCO candidates considered, one quite conclusively exhibits this behaviour while another (PN39078) hints at it in the admitting narrow TCO-supporting velocity range of a wide velocity uncertainty. Although a sample size of 1 or 2 does not support making a frequency statement, one must assume that that small sample better represents norms rather than exceptions.

Granvik et al (2012) performed an analysis of TCO capture by calculating energy states within a rotating reference frame. We did not perform an equivalent analysis. Therefore, we are unable to define the moment of capture of a candidate clone. Our apparent clone motion did not yield obvious connection to L1 and L2 as report by Granvik et al., but we cannot conclusively argue support or contradiction to their finding. However, when analysing deltas in the sum of potential and kinetics energies of clones for EN130114, we note the largest total energy drop occurs near Earth passage. This drop is assumed to be related to Earth-Moon system capture (see Figure 16). The Moon's influence of this process is not clear. It is tempting to research a dynamical trigger for the focusing of the object to Earth collision. As can be seen in Figure 6, the spread of clone anomaly along the clone cloud orbit is rapid, taking about 11 days for the clones to disperse completely around the orbit. Therefore, conclusively identifying the object's behaviour for any significant number of days prior to contact is problematic. But, as described in the results, the significant portion of TCO clones passing through the Lunar Hill Sphere on their final pre-Earth impact is compelling evidence for lunar influence.

Jedicke et al. (2014) define a dynamical class of objects called "drifters": objects which enter into geocentric motion for a brief period of time without completing one revolution within the Earth-moon system. Their simulations predict a steady state population of drifters being ten-fold that of TCOs. In the case of Earth impacting objects, the distinction between impacting drifters and meteoroids which strike the earth directly from heliocentric orbit requires definition. We did not attempt to make this distinction, and treat both scenarios simply as unbound. Our technique of counting orbits by recording transitions in Earth-object distance derivatives does appear helpful in making the distinction.

Given the low entry speed of TCOs we expect a substantial survival fraction. Our modelling of EN130114 suggests hundreds of grams of meteoric material reached the ground. In the case of JPL20140626, a simple single body ablation approach coupled with average apparent ablation coefficient of 0.014 $s^2km^{-2}$ (see Ceplecha et al., 1998) gives a survival fraction of 40% or nearly 5 tonnes. This is certainly an upper limit as fragmentation likely reduces this value substantially; nevertheless it is probably of the order of a tonne of material survived to reach the ground. The quoted peek brightness height of 28.5 km is high for a low speed, large object but we note that the event occurred in darkness and hence may represent the main fragmentation altitude as opposed to a true end height which would be more visible if the dust cloud were sunlit.

The visibility of EN130114 is of interest as an indicator to the discoverability of TCOs for exploitation. Using the Bowell et al. (1989) relationship of asteroid size to magnitude (assuming asteroids are a proxy for meteoroid visibility):

$$D = \frac{1347 \times 10^{-H_k/5}}{A_k^{1/2}}$$

the calculated diameter of 15 centimetres, and assuming an albedo $A = 0.20$, we arrive at an absolute magnitude for EN130114 of H = 36.5. Using the IAU standard asteroidal apparent magnitude calculation as documented in Bowell et al. (1989), and assuming a magnitude slope parameter of 0.15, we calculate EN130114's apparent magnitude at m = 16.7 well within the limiting magnitude of the larger sky surveys. The apparent magnitude drops quickly to m = 24.0, the approximate limiting magnitude of large surveys, within 10 hours prior to contact. However, with the large number of EN130114 probability clones within a tight Earth-centred orbit of lunar distance aphelion, and with phase angle improving over (backward) time, there is significant likelihood that EN130114 was visible periodically during previous passages of the Earth, being visible for 10-15% of its orbit period.

# 6   Conclusions

Dynamical back integrations of EN130114 using two atmospheric deceleration models lead to conclusion that:

1. EN130114 was likely a Temporarily Captured Object. The preferred deceleration models consistent with observations assuming fragmentation of a 3 g/cm$^3$ object yield a 92.1 - 98.6% likelihood; with back-integration of the large majority of statistical samples across observation measurement uncertainties yielding either TCO behaviour or the unreal scenario of an ejection from the Earth itself. A less likely 1 g/cm$^3$ fragmenting object yields a 24% likelihood. In both high and low density cases, approximately 5-8% of the TCO-consistent samples exhibit TCO behavior for up to five years prior to impact. A significant portion remains TCOs for 250-500 days.

2. Two scenarios of TCO capture from heliocentric orbit by the Earth-Moon system exist. The most likely is that the TCO entered into an elongated highly inclined orbit with apogee near the Moon's orbit, permitting the Moon to play a significant role in focusing the TCO. A less likely scenario, applicable only to the low density model is that the object entered into a broad less-inclined orbit reaching out into the outer Earth Hill Sphere, where the Sun's gravitation played a major role in TCO dynamical evolution.

3. TCOs entering into the above eccentric Earth centred orbit are detectable in current sky surveys for substantial periods of time.
4. Velocity is the all-important factor in leading to a TCO determination in this case. Radiant and positional variations within observation uncertainties have no impact on the results of clone integration.

The NASA Fireball and Bolides Report event of 20140626 has a real but small probability of being a TCO.

1. The probability ranges from 10% to 0% depending on the atmospheric deceleration model used with the published measurement uncertainties.
2. Reasonable scenarios of atmospheric deceleration eliminate the possibility of the object being a TCO. A not-unreasonable 300 m/s deceleration representing a non-fragmenting carbonaceous composition or significant meteoroid fragmentation drives the TCO probability to zero.

TCO behaviour cannot conclusively be determined from measurements reported to the precision provided by NASA Fireball and Bolides Report. Measurement precision of systems such as the Czech portion of European Network is required to conclusively determine if an observed object is a TCO. The velocity ranges consistent with an object being a TCO are narrow, 0.02 – 0.04 km/s in width in the case of EN1300114.

# 7   Acknowledgements

This work was supported in part by the NASA Meteoroid Environment Office through NASA co-operative agreement NNX15AC94A, the Natural Sciences and Engineering Research Council of Canada (NSERC) and the Canadian Space Agency (CSA) through the ASTRO CSA Cluster; and in part by Preamium Academiae from the Czech Academy of Sciences, grant

## 8 Tables and Figures

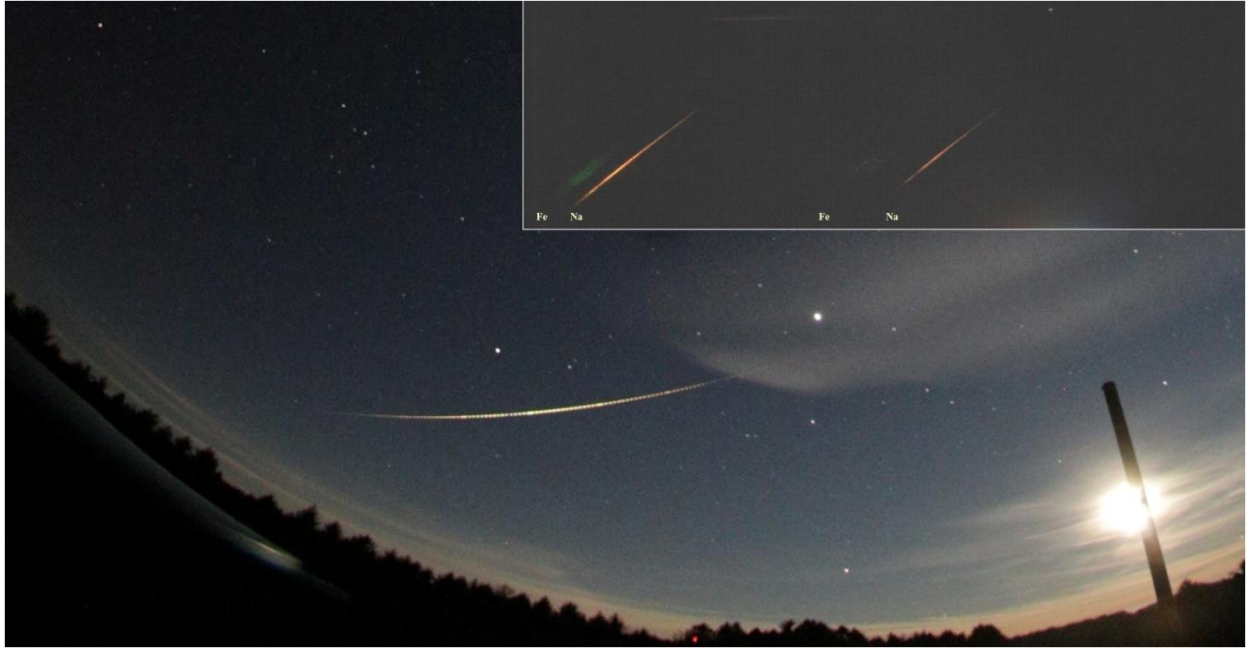

Figure 1 – Image of EN130114 event taken by the Digital autonomous observatory at Kunžak station. Inset: spectrum of EN130114 by L. Shrbený.

Table 1 - In-atmosphere trajectory, mass and fireball type determined for EN130114.

PE/Type: Empirical end height criteria and resulting fragmentation class/type, see Ceplecha & McCrosky (1976). DAFO: Digital Autonomous Fireball Observatory, DF: Digital Camera – imaging parameters are the same.

|  | Beginning | Terminal |
|---|---|---|
| Time (UT) | 3:01:37.62 UT | 3:01:45.70 |
| Height (km) | 74.589 ± 0.015 | 32.494 ± 0.009 |
| Longitude (deg E) | 13.42570 ± 0.00014 | 13.67707 ± 0.00013 |
| Latitude (deg N) | 49.07656 ± 0.00006 | 48.52321 ± 0.00005 |
| Mass (kg) | 5. | 0.2 |
| Slope (deg) | 33.302 ± 0.014 | 32.724 ± 0.014 |
| Maximum absolute magnitude | -7.6 ||
| Total length (km) / Duration (s) | 77.26/8.08 ||
| PE/Type | -4.60 / I ||
| EN stations | 02 Kunžak (DAFO), 20 Ondřejov (DF) ||

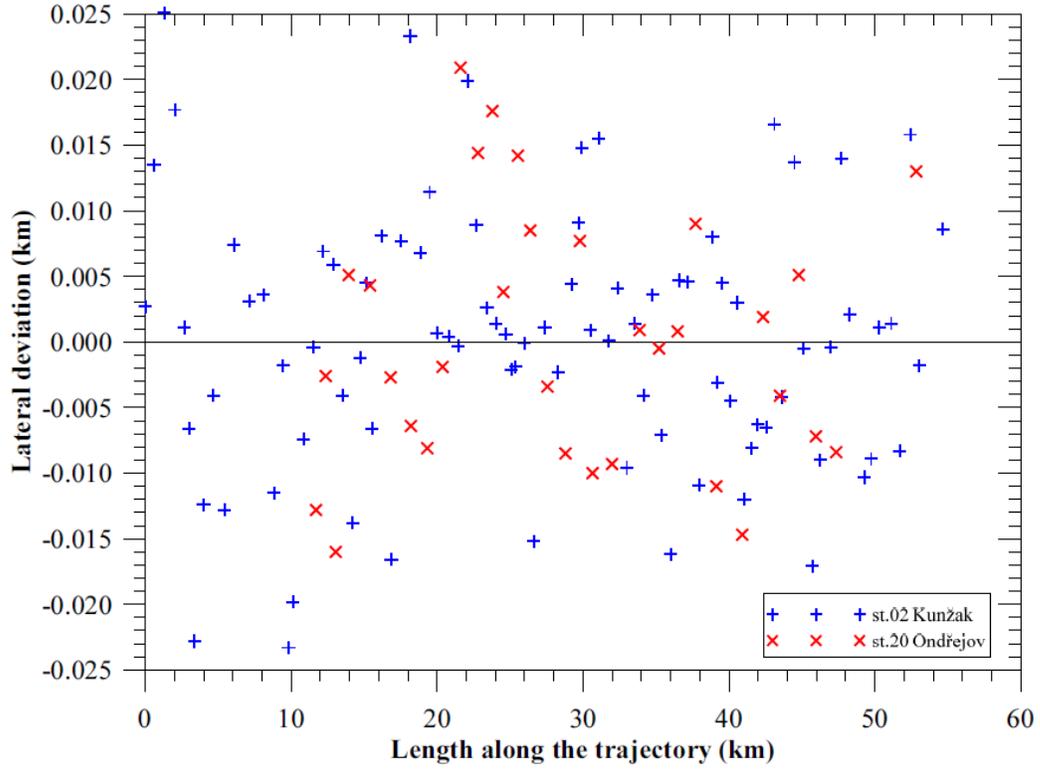

**Figure 2 - Astrometric residuals for EN130114**

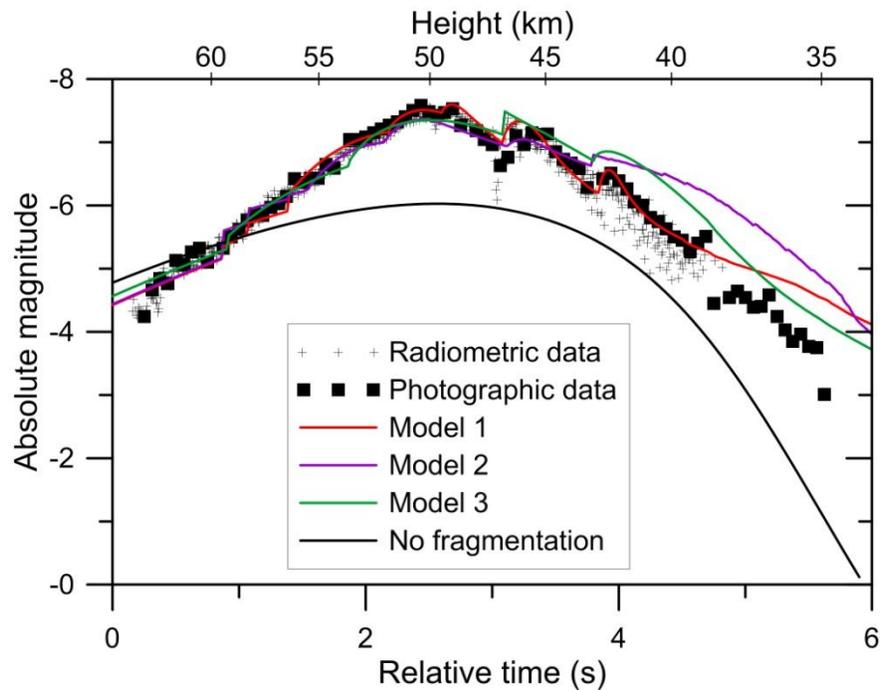

Figure 3 - EN130114 models considered comparing modelled light curves to observed photographic and radiometric curves. Models 1 and 2 correspond to upper and lower velocity extremes for a high density meteoroid scenario. Models 1 and 2 are combined for orbital integration as a High Density Model. Model 3 represents a lower density scenario representing the highest feasible initial velocity, although considered relatively unlikely (10% probability) due to the observed terminal height. A no fragmentation model was considered but does not fit the observed light curve, and therefore was discounted for further analysis.

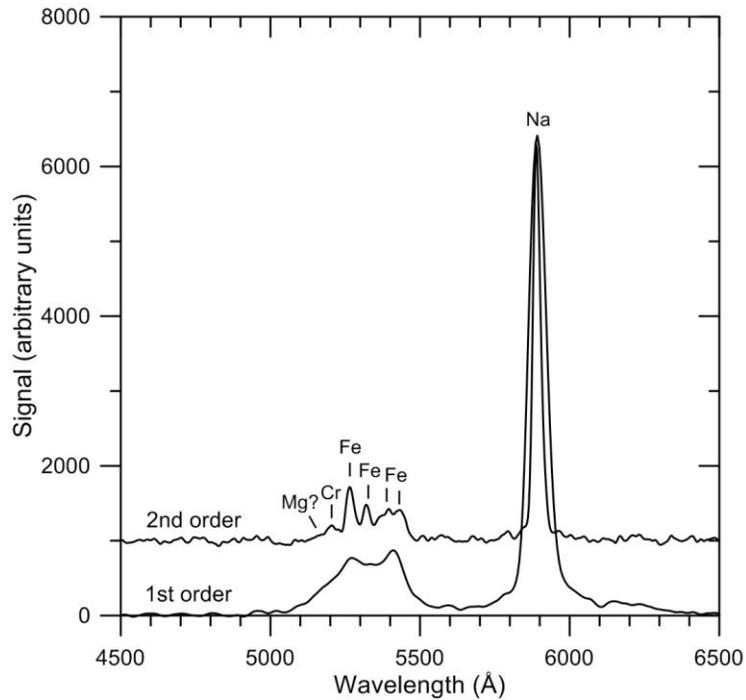

Figure 4 - Spectral plot for EN130114. The dominant emission is by neutral Na. Other visible lines belong to neutral Fe and Cr. These emissions are common in natural meteoroids (see e.g. Vojáček et al. 2015, Borovička 1994) and the spectrum is fully consistent with a natural object. The second order was offset by 1000 units for clarity. The second order has lower sensitivity but higher resolution than the 1st order.

Table 2 - The JPL20140626 and JPL200080701 events as documented in Yeomans & Baalke (2014).

The velocity components correspond to a velocity magnitude of 11.2 km/s and 9.8 km/s respectively. As stated on the site: The pre-impact velocity components are expressed in a geocentric Earth-fixed reference frame defined as follows: the z-axis is directed along the Earth's rotation axis towards the celestial north pole, the x-axis lies in the Earth's equatorial plane, directed towards the prime meridian, and the y-axis completes the right-handed coordinate system.

| Date/Time - Peak Brightness (UT) | Latitude (Deg) | Longitude (Deg) | Altitude (km) | Velocity Components (km/s) | | | Total Radiated Energy (J) | Calculated Total Impact Energy (kt) |
|---|---|---|---|---|---|---|---|---|
| | | | | vx | vy | vz | | |
| 2014-06-26 05:54:41 | 71.5S | 93.4E | 28.5 | 7.0 | 2.9 | 8.3 | 6.1E+10 | 0.2 |
| 2008-07-01 17:40:19 | 37.1N | 115.7W | 36.1 | 2.8 | 1.7 | -9.2 | 3.6E+10 | 0.12 |

Table 3 - Integration starting values and standard deviations for all events.

Angular values are all J2000; the right ascension and declination being of the apparent radiant. Cartesian velocity components are directed as in Table 2. * JPL20140626 with σ=0. 01kps is non-real; a hypothetical reduction of uncertainties demonstrating need for high precision velocity measurements. JPL20080701+1400mps is non-real, with an initial velocity increase of 1.4 k/s to characterize possible TCO behaviour.

| Event/Model | Date/Time (UT) | Lat (deg E) | Lon (deg N) | Height (km) | Radiant and Velocity | | |
|---|---|---|---|---|---|---|---|
| EN130114 | | | | | R.A. (°) | Dec (°) | v (km/s) |
| High Density Model | 2014-01-13 3:01:34.3 | +49.32555 ± 0.00006 | +13.31068 ± 0.00014 | 93.945 ± 0.005 | 35.98 ± 0.07 | 69.88 ± 0.20 | 10.917 ± 0.035 |
| Low Density Model | " | " | " | " | " | " | 11.02 ± 0.02 |
| JPL20140626 | | | | | vx (km/s) | vy (km/s) | vz (km/s) |
| Measured | 2014-06-26 05:54:41 | -71.5 ± 0.1 | +93.4 ± 0.1 | 28.54 ± 0.01 | 7.0 ± 0.1 | 2.9 ± 0.1 | 8.3 ± 0.1 |
| +100mps | " | " | " | " | 7.06 ± 0.10 | 2.93 ± 0.10 | 8.37 ± 0.10 |
| +200mps | " | " | " | " | 7.12 ± 0.10 | 2.95 ± 0.10 | 8.45 ± 0.10 |
| +300mps | " | " | " | " | 7.19 ± 0.10 | 2.99 ± 0.10 | 8.52 ± 0.10 |
| σ=0. 01kps* | " | " | " | " | 7.00 ± 0.01 | 2.90 ± 0.01 | 8.30 ± 0.01 |
| JPL20080701 | | | | | vx (km/s) | vy (km/s) | vz (km/s) |
| Measured | 2008-07-01 17:40:19 | +37.1 ± 0.1 | -115.74 ± 0.1 | 36.14 ± 0.01 | 2.8 ± 0.1 | 1.7 ± 0.1 | -9.2 ± 0.1 |
| +1400mps* | " | " | " | " | 3.2 ± 0.1 | 1.9 ± 0.1 | -10.5 ± 0.1 |
| PN39078 | | | | | R.A. (°) | Dec (°) | v (km/s) |
| Measured | 1965-11-14 07:06:14 | +41.65 ± 0.1 | -99.8 ± 0.1 | 119.85 ± 10 | 347.44 ± 1.00 | 86.575 ± 1.000 | 10.88 ± 0.10 |

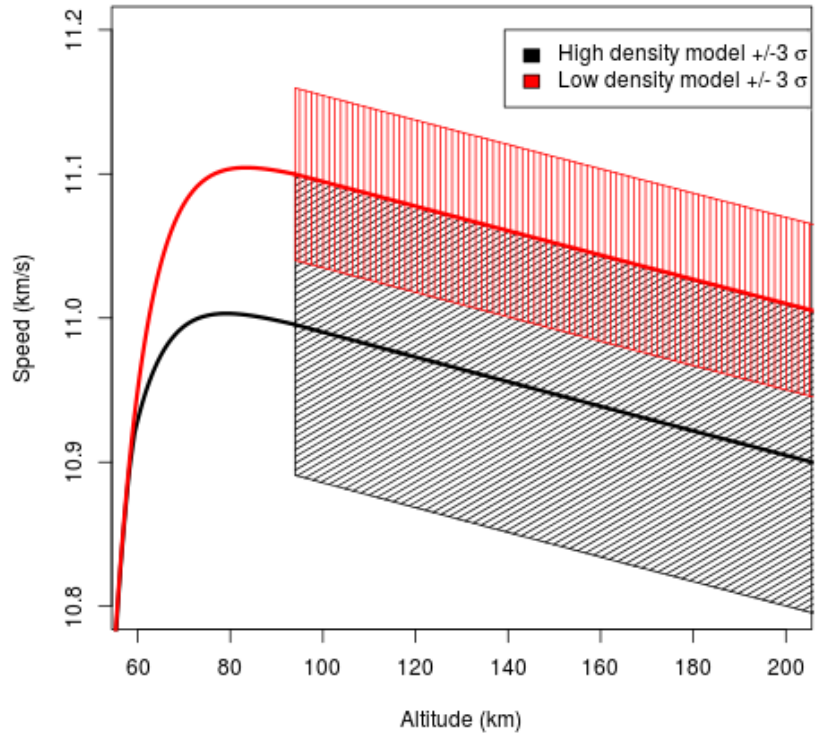

Figure 5 - Speed relative to the Earth's center versus altitude (in kilometers above the WGS84 geoid) for the EN130114 high and low density models The heavy central line indicates the nominal particle for the suite of clones, and the shaded area indicates 3 standard deviations from that value.

Table 4 - Probability clone classes used in this work.

It is understood that Hill Spheres do not represent an absolute cut-off of gravitational interactions, and that objects outside a Hill Sphere are still interacted upon by the Hill Sphere parent. Hill Spheres are strictly used to demonstrate gravitational dominance in the categorizations.

| Class | Description |
|---|---|
| TCO | Objects which exhibit TCO behaviour but do not pass through the Lunar Hill Sphere at any time |
| TCOL | Objects which exhibit TCO behaviour and interact with the Moon by passing through the Lunar Hill Sphere |
| UNB | Unbound (with Earth) objects which do not exhibit TCO behaviour and collide with Earth directly from heliocentric orbit |
| UNBL | objects which do not exhibit TCO behaviour, but travel through the lunar Hill Sphere directly prior to Earth impact |
| Sp0 | "Sputniks", physically impossible scenarios (for natural objects) where the back integrated clone is directly emitted from the Earth or Moon's surface without orbiting the Earth once |
| Spn | physically impossible scenarios of where objects are emitted from the Earth or Moon, orbiting the Earth before impacting |

Table 5 - Event probability clone behaviour by deceleration model.

The TCO Probability % is calculated as (TCO+TCOL) / (TCO+TCOL+UNB+UNBL). EN130114 clone type counts ranges and TCO probability ranges are based on pessimistic and optimistic results from two methods of Sputnik detection as described in section 3.4. Behaviour nomenclature is described in the text. A larger number of clones is used for the High Density EN130114 Model to provide >1000 TCOs to analyse. The number of JPL20140626 clones is restricted to 1000 due to low TCO likelihood. * JPL20140626 with σ=0. 01kps is non-real; a hypothetical reduction of uncertainties demonstrating need for high precision velocity measurements. JPL20080701+1400mps is non-real, with an initial velocity increase of 1.4 k/s to characterize possible TCO behaviour. ** The 37.9% TCO probability for PN39078 is highly questionable, taking into account the inability to confirm measurements, and the extreme number of impossible Sp0 & Spn clones resulting from the existing measurements.

| | | #Clones | Clone Types (see text) | | | | | | TCO |
|---|---|---|---|---|---|---|---|---|---|
| | | | TCO | TCOL | UNB | UNBL | Sp0 | Spn | Prob% |
| EN130114 High Density | Optimistic | 20000 | 804 | 2569 | 49 | 0 | 12878 | 3700 | 98.6 |
| | Pessimistic | 20000 | 61 | 508 | 49 | 0 | 12986 | 6427 | 92.1 |
| EN130114 Low Density | Optimistic | 10000 | 1486 | 438 | 6137 | 0 | 1143 | 796 | 23.9 |
| | Pessimistic | 10000 | 1392 | 415 | 6137 | 0 | 1144 | 912 | 22.7 |
| JPL20140626 | | 1000 | 54 | 9 | 537 | 0 | 391 | 9 | 10.5 |
| JPL20140626+100mps | | 1000 | 20 | 3 | 858 | 0 | 109 | 10 | 2.6 |
| JPL20140626+200mps | | 1000 | 4 | 2 | 983 | 0 | 11 | 0 | 0.6 |
| JPL20140626+300mps | | 1000 | 0 | 0 | 1000 | 0 | 0 | 0 | 0.0 |
| JPL20140626 σ=0. 01kps* | | 1000 | 228 | 42 | 652 | 0 | 26 | 52 | 29.3 |
| JPL20080701+1400mps* | | 1000 | 69 | 1 | 713 | 0 | 205 | 12 | 8.9 |
| PN39078 | | 999 | 25 | 0 | 41 | 0 | 908 | 25 | 37.9** |

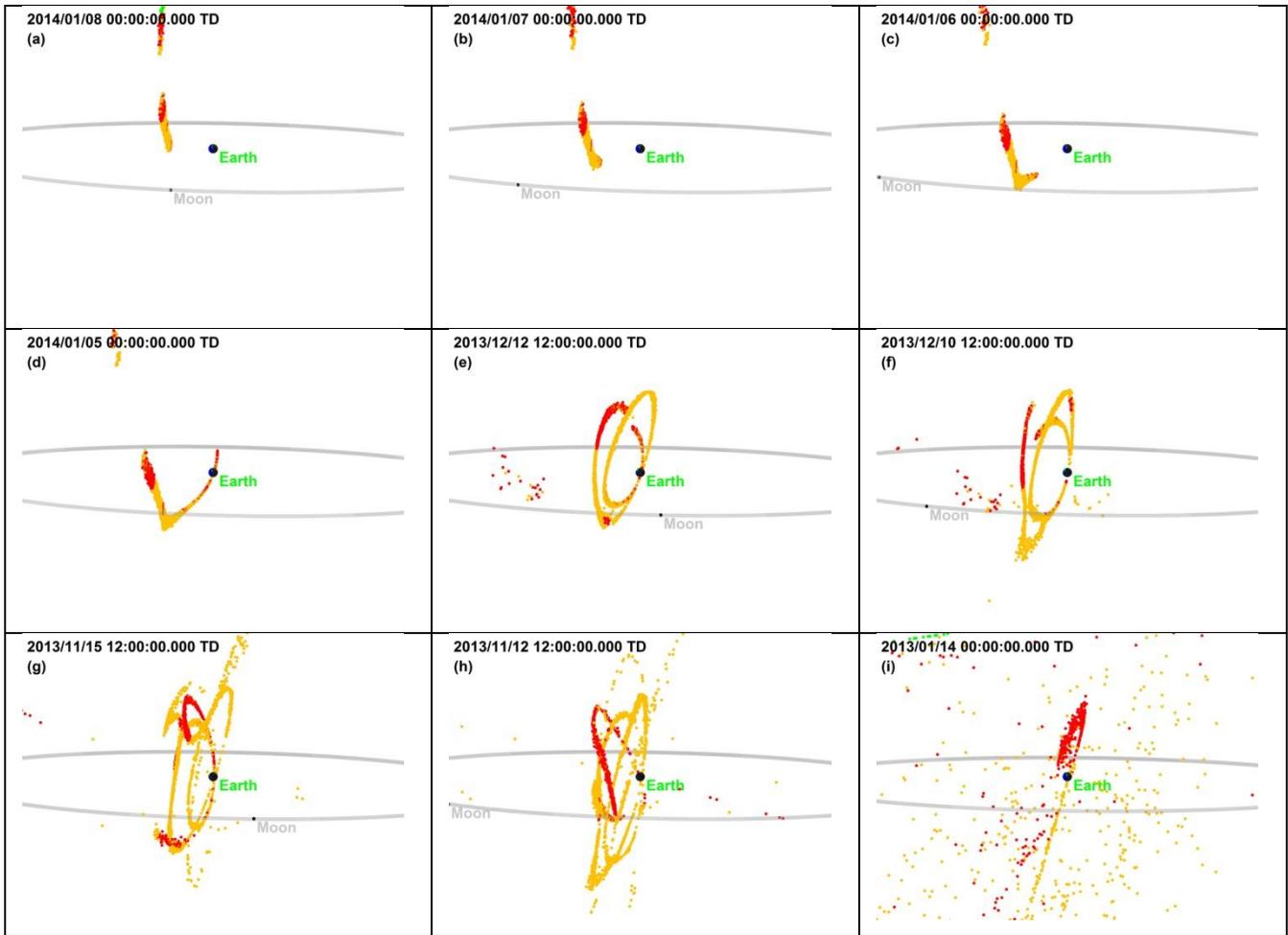

Figure 6 - EN130114 high density model low velocity probability cloud focussing by the Moon.  The Moon is orbiting right to left (in backward time, left to right in forward time) in the bottom foreground leg of its orbit.  a), b), c), d) show interaction with the first lunar passage through the clone cloud prior to the event; e), and f)  the second passage; g) and h) the third passage, and i) pre-focussed dispersion 1 year prior to the event.  Red points represent TCO clones, orange points TCOL clones.  Green are unbound.

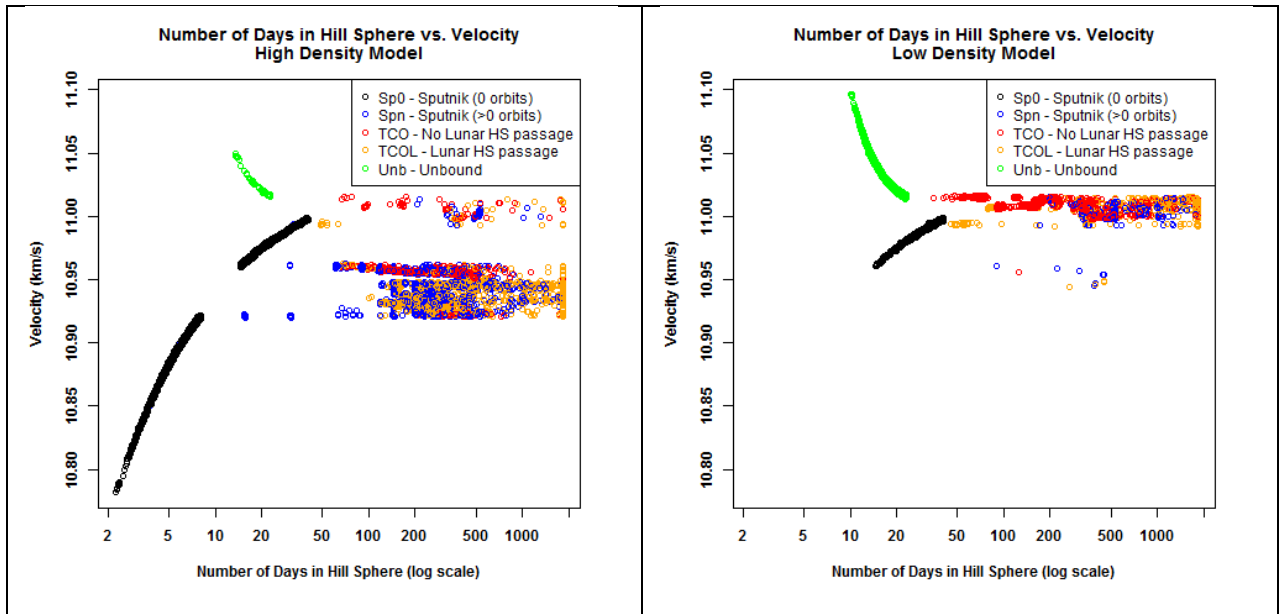

Figure 7 - Clone duration within Earth's Hill Sphere for high and low density EN130114 models by clone initial geocentric velocity, based on 20,000 clones for high density and 10,000 clones for low density. Impossible scenarios (zero orbit and >0 orbit planet emissions) are shown but disregarded for later analysis. The preferred high density model exhibits little unbound behaviour, most possible scenarios are TCO, with a majority of TCO's showing lunar Hill Sphere interaction. The less likely low density model exhibits some TCO behaviour for the lowest possible velocities. Two TCO velocity bands are evident, a lower velocity band most prominent in the high density model, and a higher velocity band prominent in the low density model. The low density low velocity band appears limited by observed velocity only.

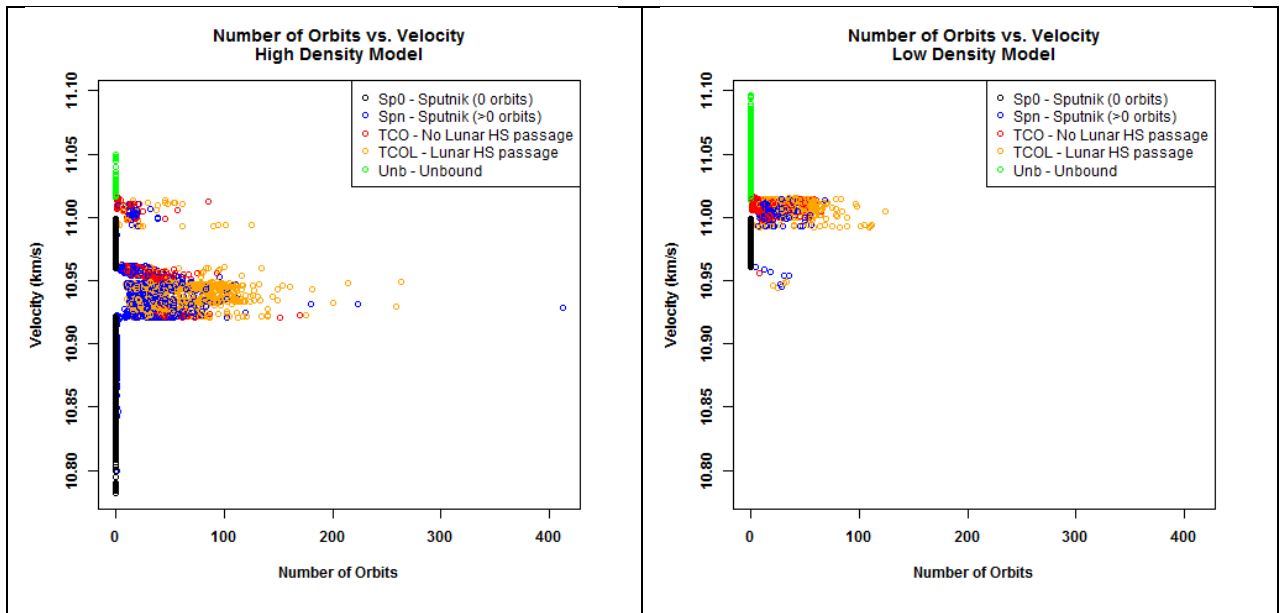

Figure 8 – The number of clone orbits for each EN130114 model corresponding to the Earth's Hill Sphere plots of Figure 7.

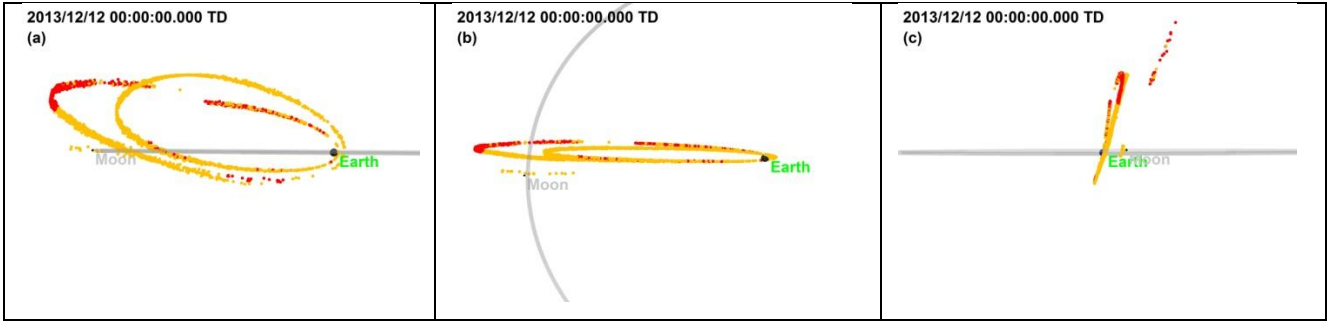

Figure 9 - EN130114 high density model low velocity TCO clone cloud orbital orientation with respect to the lunar orbit (grey).  a) View from the plane of Moon's orbit showing fullest extent of the clone orbits.  b) View from above the Moon's orbit.  c) View from the plane of the Moon's orbit showing inclination of the cloud.

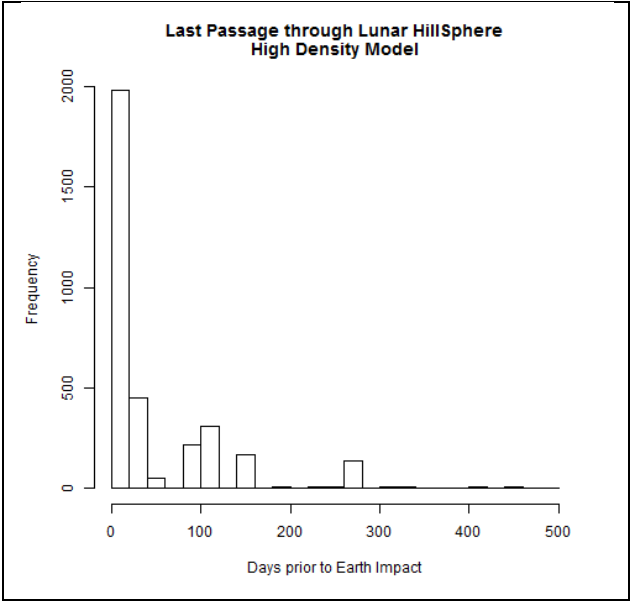

Figure 10 –Histogram of EN130114 clone last passage through the lunar Hill Sphere prior to Earth impact, the large majority of which pass through the lunar Hill Sphere immediately before hitting Earth.

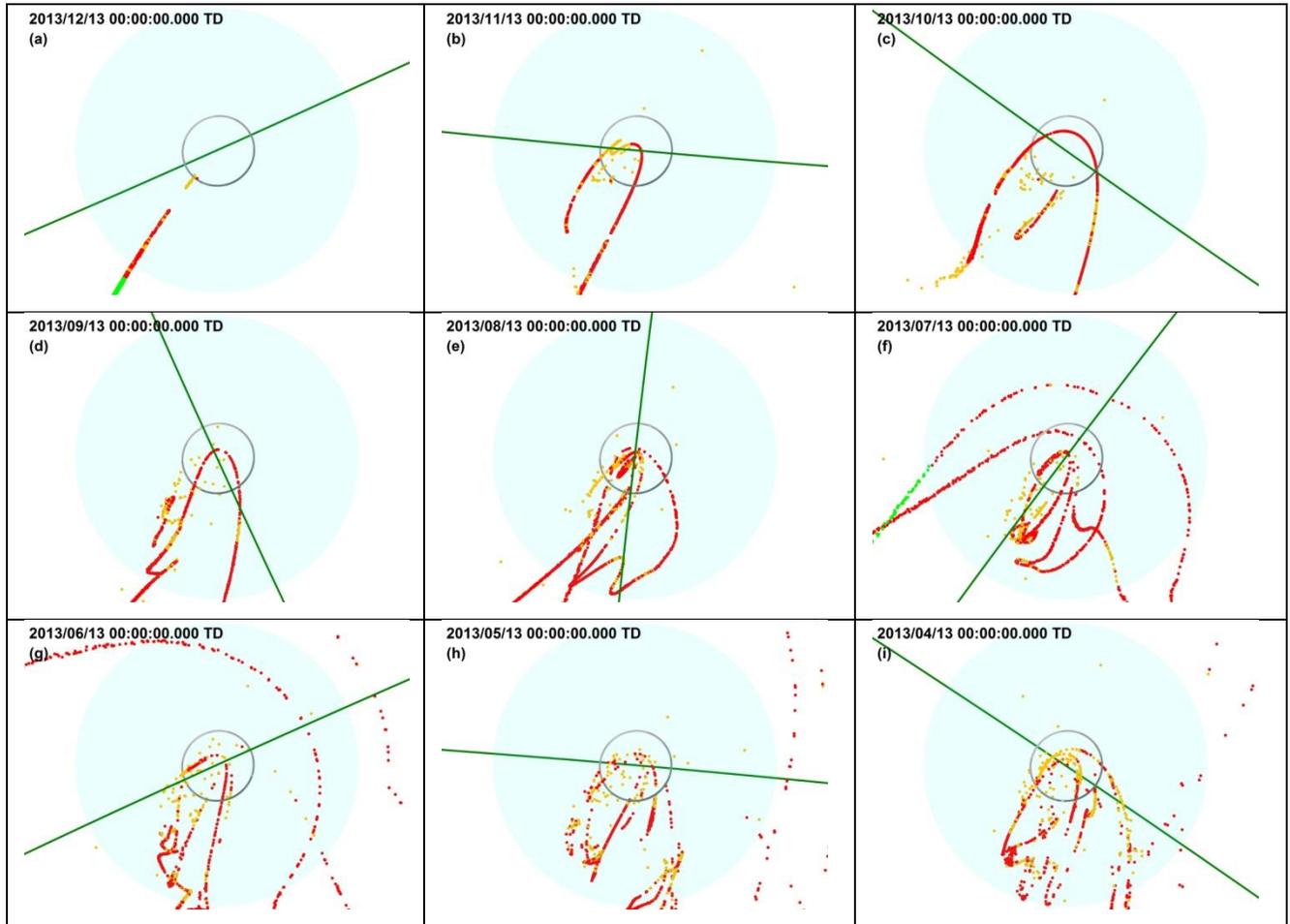

Figure 11 - EN130114 low density model high velocity clone evolution over 9 months at 1 month intervals. With clone back-trajectories reaching well out into the Earth's Hill Sphere (blue disk), the Sun's gravity influence is greater with the Sun playing the greater role in clone evolution. Red points represent TCO clones, orange points TCOL clones. Green are unbound. The Moon's orbit is in grey, the Earth's orbit in green.

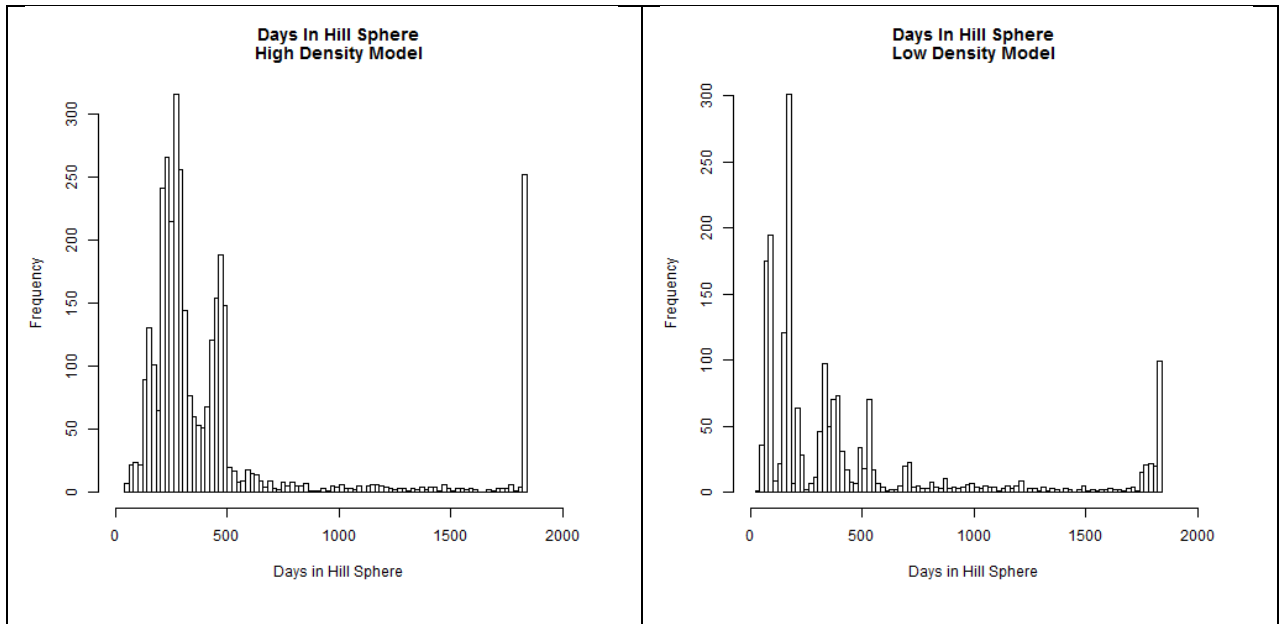

Figure 12 - EN130114 frequency plots of the number of days clones remained with the Earth Hill Sphere. The frequency spikes at the end of the graphs represent clones still in orbit at the end of a 5-year integration. The high density graph applies to a population of 3422 physically possible clones, 3373 of which are TCOs (See Table 5). The low density graph applies to 8061 physically possible clones, 1924 of which are TCOs.

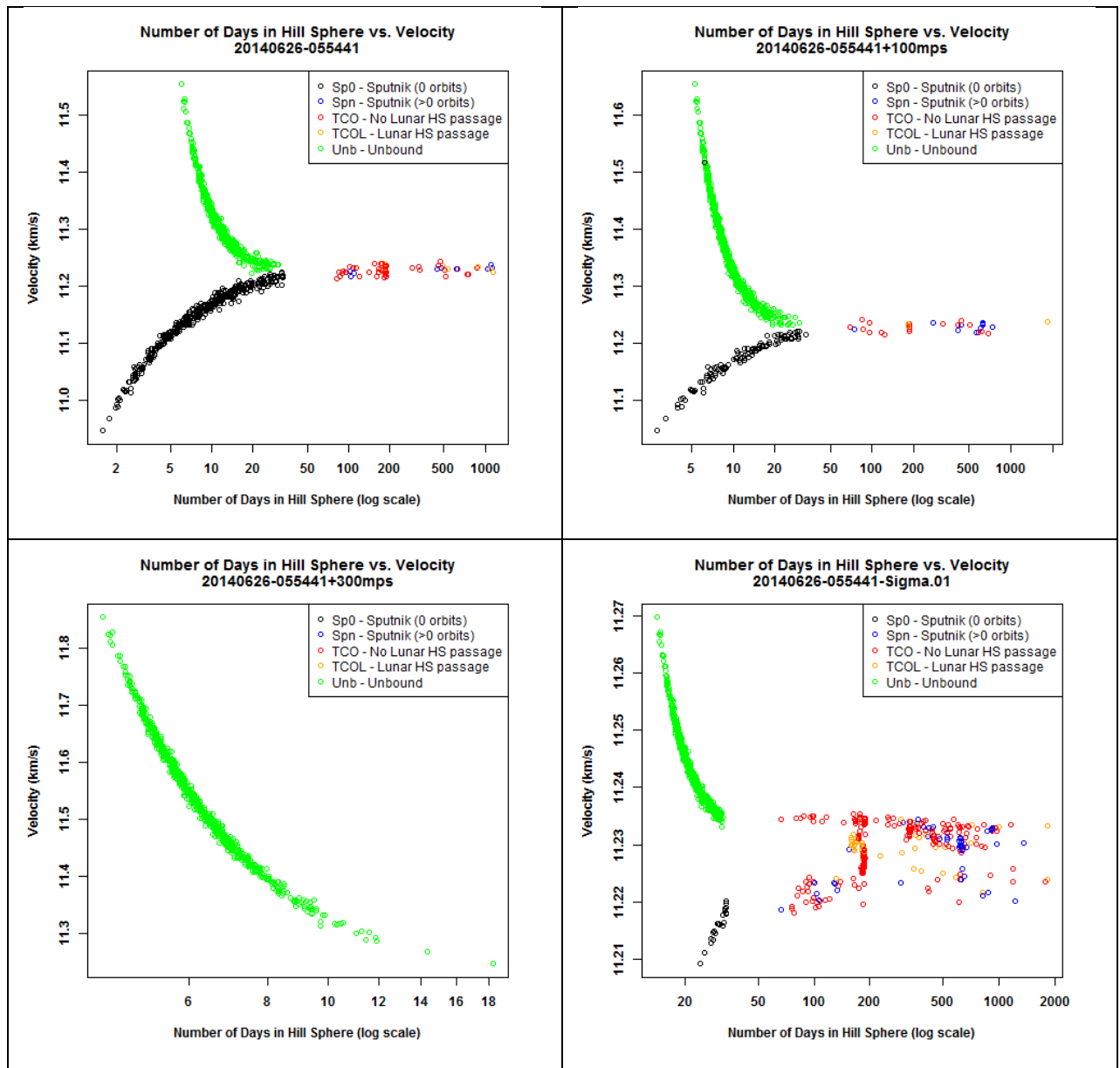

**Figure 13** - Clone duration within Earth's Hill Sphere for JPL20140626 assuming measured velocity (no deceleration), atmospheric deceleration of 100 m/s and 300 m/s (200 mps omitted for brevity), and the hypothetical scenario reported uncertainties are reduced from 0.1 km/s to .01 km/s. All TCO activity ceases with 300 m/s deceleration assumed. The 0.01 sigma plot indicates the need for high precision measurements of velocity in order to conclude TCO behaviour.

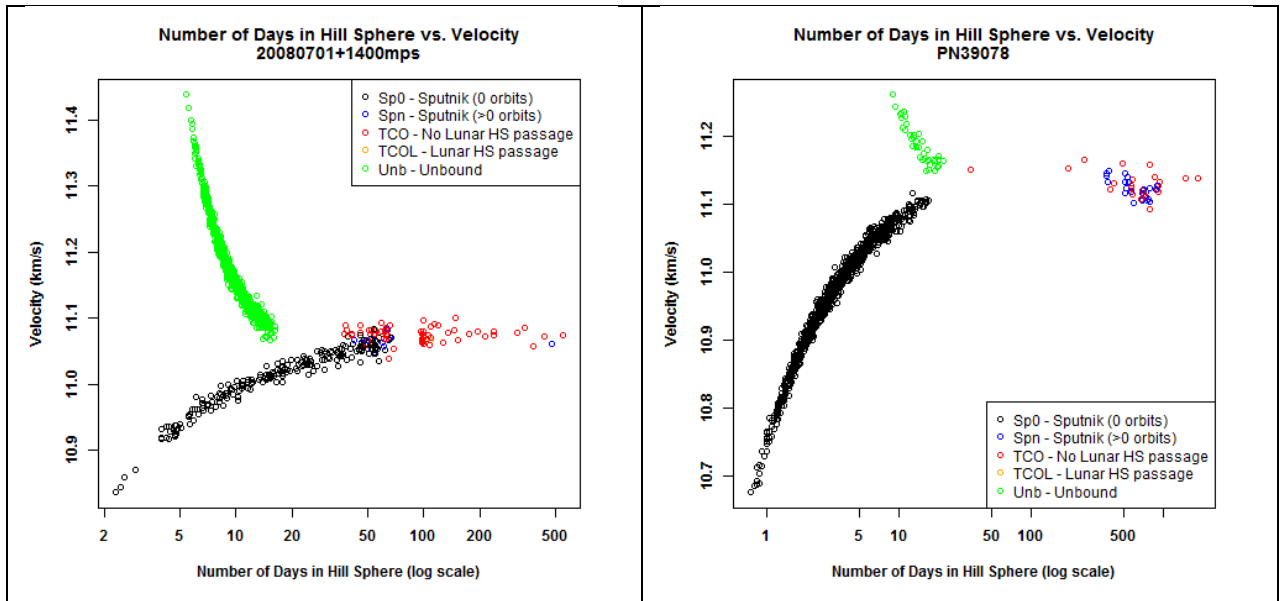

Figure 14- Left: Clone durations with Earth's Hill Sphere for artificially accelerated JPL20080701 and for PN39078. Velocity uncertainties are too large to make statements on the events being TCO's other than that small possibility exists. More precise velocity measures are required.

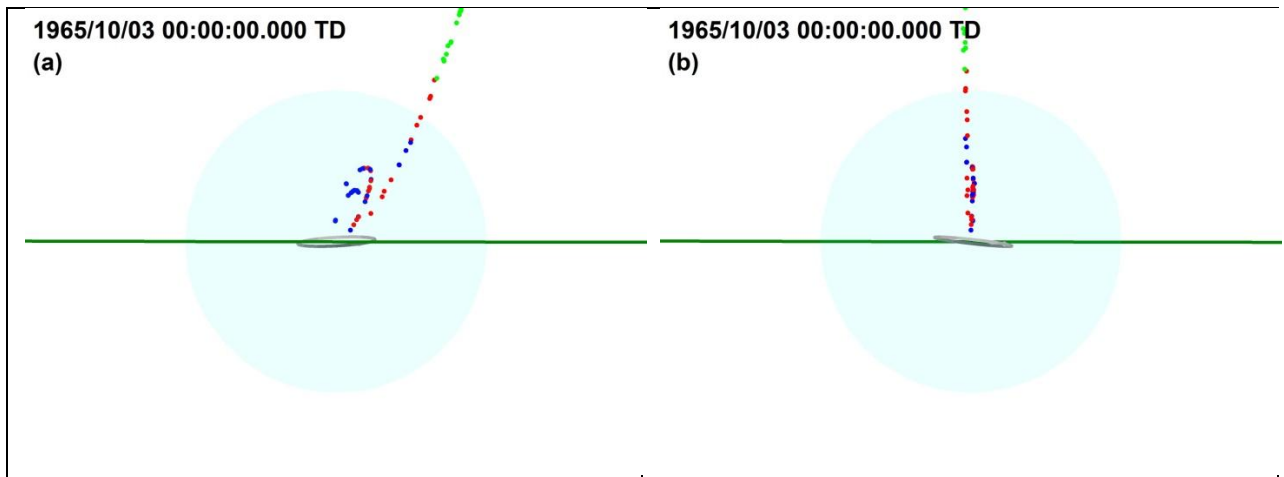

Figure 15 - PN39078 TCO and Spn clones in a highly inclined orbits, some of which resemble the highly inclined lunar distance orbit of EN130114. Views are 90° apart from the Earth's ecliptic. The Earth's orbit is in green, the Moon in grey. TCO clones are in red, Spn blue, and unbound green. The blue disc is the Earth's Hill Sphere.

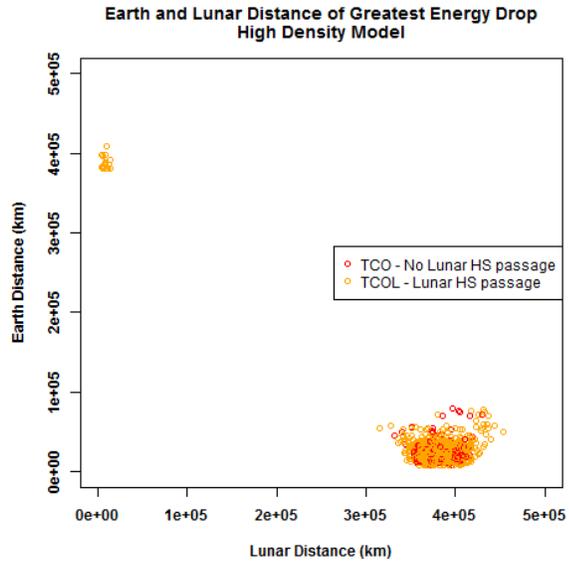

**Figure 16 -** Clone distances from the Earth and Moon at the point of largest total energy drop, assumed to be the point of capture by the Earth-Moon system. Total energy is the sum of the kinetic energy with respect to the Earth-Moon barycentre, the potential energy with respect to Earth and the potential energy with respect to the Moon.